\documentclass{iopart}
\usepackage{graphicx}

\def\Journal#1#2#3#4{{#1} {\bf #2}, #3 (#4)}

\def\NPB{Nucl. Phys. B}
\def\PLB{Phys. Lett. B}

\def\PRD{Phys. Rev. D}

\def\EPJ{Eur. Phys. J. C}

\newcommand{\ra}{\rightarrow}
\newcommand{\D}{D}
\newcommand{\Dbar}{\bar{D}}
\def\lsim{\mathrel{\rlap{\lower4pt\hbox{\hskip1pt$\sim$}}
    \raise1pt\hbox{$<$}}}

\begin{document}

\title{Polarized Parton Distribution in Neutrino Induced 
Heavy Flavor Production}

\author{Kazutaka SUDOH}

\address{Radiation Laboratory, 
RIKEN (The Institute of Physical and Chemical Research), 
Wako, Saitama 351-0198, JAPAN \\
E-mail: {\tt sudou@rarfaxp.riken.go.jp}}

\begin{abstract}
In order to examine polarized strange quark distribution, semi--inclusive 
$\D$/$\Dbar$ production in neutrino deep inelastic scattering 
is studied including ${\cal O}(\alpha_s)$ corrections.
Cross section and spin asymmetry are calculated by using various 
parametrizations of polarized parton distribution functions.
It is found that $\Dbar$ production is promising to directly extract 
the polarized strange sea. 
\end{abstract}



%\maketitle

\section{Introduction}
A study of heavy flavor production in deep inelastic scattering (DIS) is 
one of the most promising ways to access the parton density in the nucleon.
As is well known, the polarized parton distribution function (PDF) plays 
an important role in deep understandings of spin structure of the nucleon.
However, knowledge about the polarized sea--quark and gluon distributions 
remain still poor, and theoretical and experimental ambiguities are rather 
large. 
In particular, flavor structure of the sea--quark distribution has been 
actively studied in these years.
There are so far several parametrization models of sea--quark distributions.
Though the simplest case is to assume the flavor SU(3)$_f$ symmetry, 
a new parametrization including the violation of the SU(3)$_f$ 
symmetrty has been also proposed \cite{Leader99}.
In order to understand the spin structure of the nucleon, we need more 
information about the polarized sea--quark and gluon distribution functions.

Charged current (CC) DIS is effective to extract the flavor decomposed 
polarized PDFs, since $W^{\pm}$ boson changes the flavor of parton.
Since there is no intrinsic heavy flavor component in the nucleon, 
we can extract information of the parton flavor in the nucleon from 
the study of heavy flavor production in CC DIS.
Actually, the NuTeV collaboration reported a measurement of unpolarized 
$s$ and $\bar{s}$ quark distributions by measuring dimuon cross sections 
in neutrino-DIS \cite{NuTeV01}.

In this work, to extract information about the polarized PDFs we 
investigated $\D$/$\Dbar$ meson production in CC DIS including ${\cal O}
(\alpha_{s})$ corrections in neutrino and polarized proton scattering; 
$\nu + \vec{p} \ra l^- +D+X$, $\bar{\nu} + \vec{p} \ra l^+ + \Dbar + X$.
The leading order process is due to $W$ boson exchange $W^+ s(d)\ra c$.
In addition, several processes are taken account of ${\cal O}
(\alpha_{s})$ next-to-leading order (NLO) calculations, in which gluon 
radiation processes $W^+ s(d) \ra cg$, virtual corrections to remove 
singularity coming from soft gluon radiation, and boson--gluon fusion 
processes $W^+ g\ra c\bar{s}(\bar{d})$ are considered.
These processes might be observed in the forthcoming neutrino experiments.

\section{Charm Production in CC DIS}
We have numerically calculated the spin-dependent cross section and the spin 
asymmetry $A^D$ which is defined by
\begin{equation}
A^D \equiv \frac{d\sigma (-)/dx - d\sigma (+)/dx}
{d\sigma (-)/dx + d\sigma (+)/dx}
=\frac{d\Delta\sigma /dx}{d\sigma /dx} ,
\end{equation}
where $+$ and $-$ denote the helicity of the target proton.
The spin-dependent cross section can be written in terms of polarized 
structure functions $g_i$ as follows:
\begin{equation}
\fl
\frac{d^3 \Delta\sigma^{\nu p}}{dxdydz}=\frac{G_F^2 s}{2\pi (1+Q^2/M_W^2)^2}
\left[ (1-y) g_4^{W^{\mp}} + y^2 x g_3^{W^{\mp}} 
\pm y(1-\frac{y}{2})x g_1^{W^{\mp}} \right] .
\end{equation}
The $+$ and $-$ in front of the 3rd term correspond to 
when initial beam is anti-neutrino and neutrino, respectively.
Kinematical variables $x$ and $y$ are Bjorken scaling variable and 
inelasticity, respectively, defined according to the standard DIS 
kinematics, and $z$ is defined 
by $z=P_p \cdot P_D/P_p \cdot q$ with $P_p$, $P_D$ and $q$ being the 
momentum of proton, $D$ meson, and photon, respectively.
The polarized structure functions $g_i$ in $\nu \vec{p}$ scattering are 
obtained by the following convolutions:
\begin{eqnarray}
\fl
{\cal G}_i^c (x,z,Q^2) = \Delta s' (\xi,\mu^2_F)D_c(z) \nonumber \\
+
\frac{\alpha_s(\mu^2_R)}{2\pi}
\int_{\xi}^1 \frac{d\xi '}{\xi '} 
\int_{\max(z,\zeta_{\min})}^1 \frac{d\zeta}{\zeta} 
\left\{ \Delta H_i^q (\xi ',\zeta,\mu^2_F,\lambda) 
\Delta s' (\frac{\xi}{\xi '},\mu^2_F) \right. \nonumber \\
+
\left. \Delta H_i^g (\xi ',\zeta,\mu^2_F,\lambda) 
\Delta g(\frac{\xi}{\xi '},\mu^2_F)
\right\}
D_c(\frac{z}{\zeta}) ,
\end{eqnarray}
where $\Delta s'$ means $\Delta s' \equiv |V_{cs}|^2 \Delta s + 
|V_{cd}|^2 \Delta d$ with CKM parameters.
$\Delta H_i^{q, g}$ are coefficient functions of quarks and gluons, which can 
be calculated by using perturbative QCD.
$D_c (z)$ represents the fragmentation function of an 
outgoing charm quark decaying 
to $D$ meson, and we adopted the parametrization proposed by Peterson 
{\it et al.} \cite{Peterson83}.
${\cal G}_i$ is related to the polarized structure functions through 
${\cal G}_1 \equiv g_1 /2$, ${\cal G}_3 \equiv g_3$, and 
${\cal G}_4 \equiv g_4 /2\xi$.
Similar analyses have been done by Kretzer {\it et al.}, 
in which charged current charm production at NLO in $ep$ and $\nu p$ 
scattering is discussed \cite{Kretzer97}.

\section{Numerical Results}
In numerical calculations, we used the GRV98\cite{GRV98} and 
MRST99\cite{MRST99} parametrizations as the unpolarized PDFs.
As for the polarized PDFs, we adopted the AAC00\cite{AAC00}, BB02\cite{BB02}, 
GRSV01\cite{GRSV01}, and LSS02\cite{LSS02} parametrizations which are 
now widely used.

\begin{figure}[t!]
   \includegraphics[scale=0.265,clip]{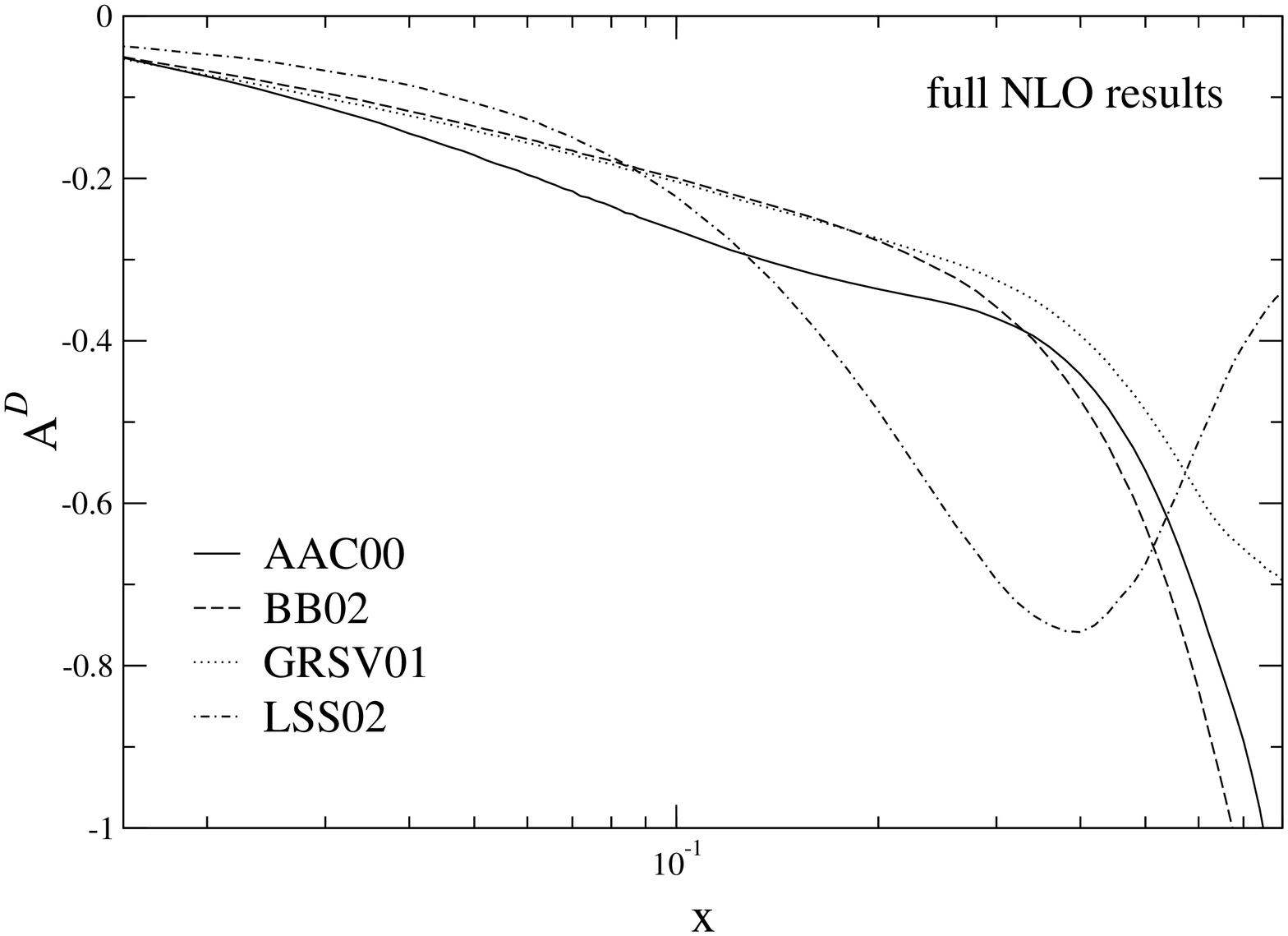}
\hspace{0.2cm}
   \includegraphics[scale=0.265,clip]{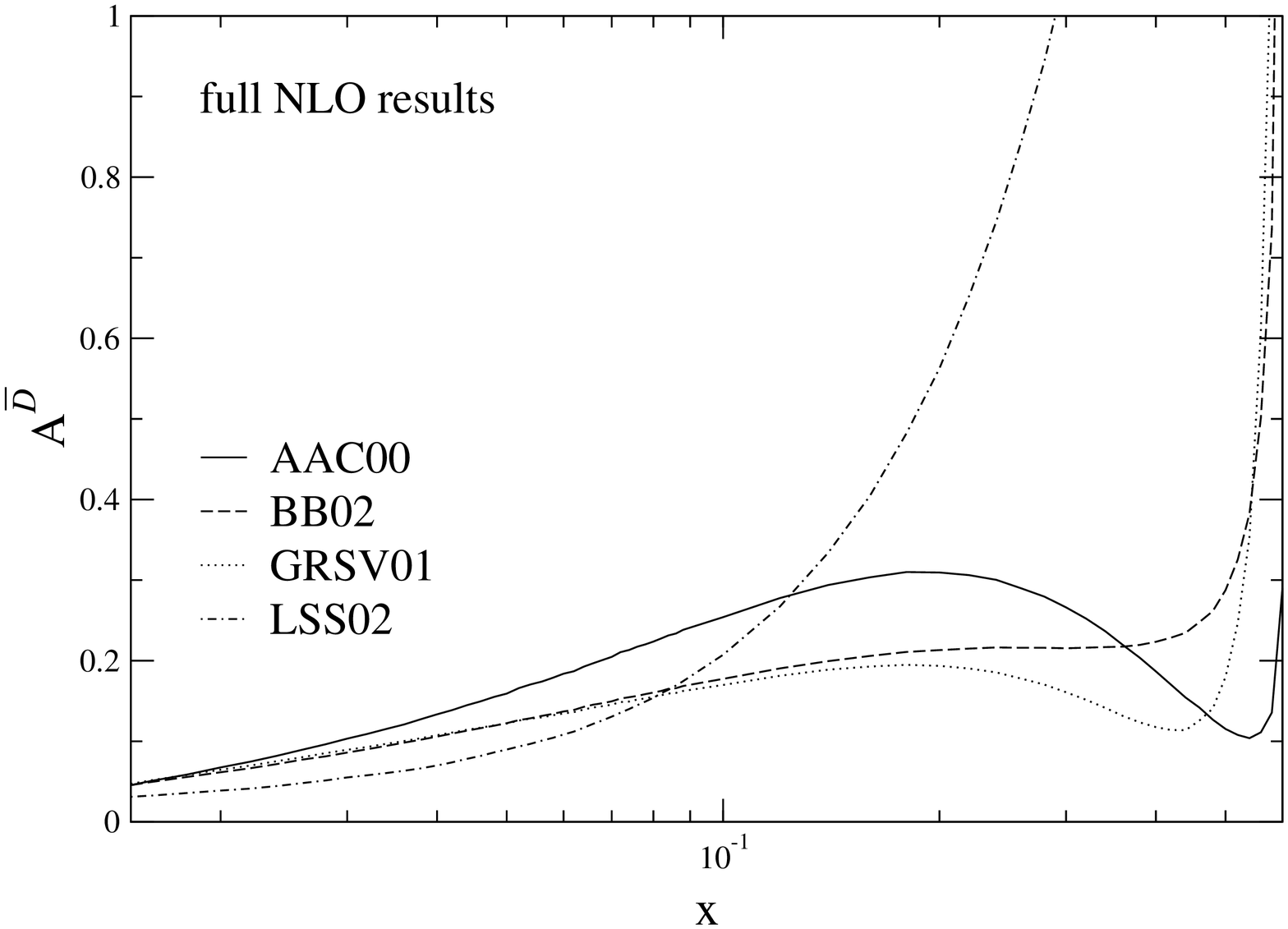}
\caption{$x$ distribution of spin asymmetries in NLO at $E_{\nu}=200$ GeV 
for $\D$ production (left panel) and $\Dbar$ production (right panel).
Solid, dashed, dotted, and dot-dashed lines show the case of AAC00 set-2, 
BB02 scenario-2, GRSV01 standard set, and LSS02 $\overline{\mbox{MS}}$ 
parametrizations, respectively.}
\label{fig}
\end{figure}

We show the spin asymmetry $A^D$ in Fig. \ref{fig} as a function of $x$ 
at initial neutrino beam energy $E_{\nu}=200$ GeV. 
Left panel and right panel in Fig. \ref{fig} represent asymmetries for 
$\D$ production and $\Dbar$ production, respectively.
For $\D$ production, $s$, $d$ quarks and gluon distribution contribute 
to the asymmetry $A^D$.
$A^D$ is dominated by valence $d_v$ quark at large $x$ ($x > 0.3$), 
though the $d$ quark component is quite highly suppressed by CKM.
On the contrary, for $\Dbar$ production, $\bar{s}$, $\bar{d}$ quarks 
and gluon component contribute to the asymmetry $A^D$.
The $\bar{d}$ quark contribution is almost negligible.
Therefore, the asymmetry is directly affected by the shape of the $\bar{s}$ 
quark distribution.
As shown in both figures, spin asymmetries strongly depend on parametrization 
models.  
We see that the case of the LSS parametrization is quite different from 
the ones of other parametrizations.
In particular, the asymmetry by the LSS parametrization in $\Dbar$ production 
goes over 1 at $x\sim 0.3$, though the asymmetry should be less than 1.
This is because the polarized $s$ quark distribution in their parametrization
extremely violates the positivity condition at $x\sim 0.3$. 
%
%NLO asymmetries by GRSV and BB parametrizations are quite similar in 
%appropriate $x$ regions ($x \lsim 0.2$), though we see some differences 
%in LO asymmetries.
%
Measurement of $\Dbar$ production in this reaction is effective 
to test the parametrization models of the polarized PDFs.
In semi-inclusive DIS, we have an additional ambiguity coming from the 
fragmentation function.
However, the ambiguity can be neglected in the $x$ distribution
of $A^D$, since the kinematical variable related to fragmentation 
is integrated out in this distribution.

\section{Summary}
In summary, semi-inclusive $\D$/$\Dbar$ meson production in CC DIS in 
neutrino--polarized proton scattering is discussed.
The cross sections and the spin asymmetries are calculated including ${\cal O}
(\alpha_{s})$ corrections with various parametrization models of polarized 
PDFs.
The $\Dbar$ production is promising to examine the sea--quark density. 
This is not the case for the $\D$ production because of the large 
$d_v$ contribution over the sea--quark contribution.
If the gluon polarization $\Delta g(x, Q^2)$ is fixed by RHIC experiments 
with high accuracy, we can directly extract the strange sea $\Delta s(x, Q^2)$. 

\end{document}